\documentclass[a4paper]{jpconf}
\usepackage{graphicx,calc,epsfig,pstricks,bbm} 
\usepackage{amsmath,amssymb,amsfonts} 
\newcommand{\be}{\begin{equation}}
\newcommand{\ee}{\end{equation}}
\newcommand{\scr}{\scriptscriptstyle}
\newcommand{\tetraedro}{{\begin{array} {c}
\ifx\JPicScale\undefined\def\JPicScale{1}\fi
\psset{unit=\JPicScale mm}
\psset{linewidth=0.3,dotsep=1,hatchwidth=0.3,hatchsep=1.5,shadowsize=1,dimen=middle}
\psset{dotsize=0.7 2.5,dotscale=1 1,fillcolor=black}
\psset{arrowsize=1 2,arrowlength=1,arrowinset=0.25,tbarsize=0.7 5,bracketlength=0.15,rbracketlength=0.15}
\begin{pspicture}(2,0)(3.33,3.33)
\psline[linewidth=0.15](0.67,0.67)(3.33,0.67)
\psline[linewidth=0.15](0.67,0.67)(2,1.33)
\psline[linewidth=0.15](2,1.33)(3.33,0.67)
\psline[linewidth=0.15](0.67,0.67)(2,2.67)
\psline[linewidth=0.15](2,2.67)(3.33,0.67)
\psline[linewidth=0.15](2,1.33)(2,2.67)
\psline[linewidth=0.15](2,2.67)(2,3.33)
\end{pspicture}
\end{array}\!\!\!}}

\newcommand{\triangolo}{{\begin{array}{c}
\ifx\JPicScale\undefined\def\JPicScale{1}\fi
\psset{unit=\JPicScale mm}
\psset{linewidth=0.3,dotsep=1,hatchwidth=0.3,hatchsep=1.5,shadowsize=1,dimen=middle}
\psset{dotsize=0.7 2.5,dotscale=1 1,fillcolor=black}
\psset{arrowsize=1 2,arrowlength=1,arrowinset=0.25,tbarsize=0.7 5,bracketlength=0.15,rbracketlength=0.15}
\begin{pspicture}(2,0)(3.33,3.33)
\psline[linewidth=0.15](0.67,0.67)(3.33,0.67)
\psline[linewidth=0.15](0.67,0.67)(2,2.67)
\psline[linewidth=0.15](2,2.67)(3.33,0.67)
\psline[linewidth=0.15](2,2.67)(2,3.33)
\end{pspicture}	
\end{array}\!\!\!}}

\begin{document}
\title{Regularized Hamiltonians and Spinfoams}

\author{Emanuele Alesci}

\address{
Universit\"at Erlangen, Institut f\"ur Theoretische Physik III, Lehrstuhl f\"ur Quantengravitation Staudtstrasse 7, D-91058 Erlangen, EU}

\ead{alesci@theorie3.physik.uni-erlangen.de}

\begin{abstract}
We review a recent proposal for the regularization of the scalar constraint of General Relativity in the context of LQG. The resulting constraint presents strengths and weaknesses compared to Thiemann's prescription. The main improvement is that it can generate the 1-4 Pachner moves and its matrix elements contain $15j$ Wigner symbols, it is therefore compatible with the spinfoam formalism: the drawback is that Thiemann anomaly free proof is spoiled because the nodes that the constraint creates have volume. \end{abstract}

\section{Introduction}
The relation between Canonical \cite{lqgcan} and Covariant \cite{lqgcov} loop quantum gravity  (LQG) \cite{lqg} is still one of the main open issues of this approach to Quantum Gravity. The beauty of the first is that provides an anomaly free quantization of the Dirac algebra but it's difficult to extract the Physical Hilbert space from the dynamics; this problem is bypassed by the spinfoam formalism, where using the quantization of a constrained BF theory, one directly defines transition amplitudes \cite{scattering,lqg} using a vertex expansion of the spinfoam model. However these objects are defined for fixed complexes and is unclear how to recover the full diff invariant theory. Nevertheless these two approaches should describe the same theory but, in spite of the fact that the spinfoam formalism was originally conceived as an exponentiation of the canonical evolution \cite{ReisenbergerRovelli97}, aimed to implement a Projector on the physical Hilbert Space, it is still not possible to formally relate the dynamics of the two, (see \cite{Io karim e francesco,me2}).
Remarkably however in 3-d, the relation between the two formalisms has been clarified in \cite{Noui Perez} and in 4-d, the spinfoam 
 \emph{kinematics} matches the canonical one \cite{Engle:2007uq,Kaminski:2009fm}.  
One thus hopes to link the two approaches  starting from the common kinematics as has been done in 3-d: Given a boundary state, the amplitude for a transition is given by a sum over histories of spinnetworks, based on 2-complexes, compatible with that state. This sequence is generated by elementary evolutions of spinnetworks taking places at the vertices of the 2-complex.  To relate the dynamics one can then proceed looking at the Hamiltonian as the generator of these evolutions.
Thiemann's Hamiltonian constraint \cite{Thiemann96a,Thiemann96b} is the natural candidate for this action, however acting on spinnetworks nodes it creates new ``extraordinary links" joined to the original graph by ``extraordinary nodes" without volume (Thiemann construction to this aim uses the AL version of the volume \cite{AshtekarLewand98}): this feature, fundamental for the consistent quantization of the Dirac algebra, is nevertheless incompatible with the spinfoam description that in 4d, on a simplicial setting, requires a vertex amplitude based on a bulk 4-simplex: this implies that the Hamiltonian acting on a 4-valent node (dual to a tetrahedron) should produce four resulting 4-valent nodes: in other words it should be able to implement a 1-4 Pachner move. The natural question is then if it's possible to find an Hamiltonian that does; the answer is yes, and an example is a suitable modification of Thiemann's one appeared in \cite{me1}.  

\section{Thiemann constraint and the new proposal}

Here we consider only the ``euclidean"  hamiltonian constraint ${\cal H}=-2{\rm Tr}[F \wedge e]$.   Following Thiemann \cite{Thiemann96a}, and choosing units where $8\pi Gc^{-3}\gamma=1$, we can write 
  $e_a^i(x) = \{A_a^i(x),V\} $, 
where $V$ is the volume of an arbitrary region $\Sigma$ containing the point $x$. Using this, and smearing the constraint with a lapse function $N(x)$, we have 
$
  \mathcal{H}[N] = 
       \int_\Sigma d^3x \, N(x)\, \mathcal{H}(x)=-2 \int_\Sigma  
      \, N  \ {\rm Tr}(F \wedge 
       \{ A, V \})~\nonumber
$.
In order to regularize this expression we review two options: 

a) {\it Thiemann construction} \cite{Thiemann96a,Thiemann96b} based a triangulation $T$ of the manifold $\Sigma$ into elementary tetrahedra with analytic edges. In this case we
take a tetrahedron $\Delta$ of $T$, and a vertex $v$ of this tetrahedron. Call the three edges that meet at $v$ as $s_i$, $i=1,2,3$ and denote $a_{ij}$ the edge connecting the two end-points of $s_{i}$ and $s_{j}$ opposite to $v$; in this way $s_{i}$, $s_{j}$ and $a_{ij}$ form a triangle.  Denote this triangle as $\alpha_{ij} := s_{i} \circ a_{ij} \circ
s_j^{-1}$. 
Decompose the smeared Euclidean constraint into a sum of one term per each tetrahedron
\begin{eqnarray}
  \label{Ham_T1}
  \mathcal{H}[N] = \sum_{\Delta \in T}
        {-2}\, \int_\Delta d^3x \, N 
            \ \epsilon^{abc}\ {\rm Tr}(F_{ab}
            \{ A_c, V \}) ~.
          \label{Ham_T2}
\end{eqnarray}
Define the 
classical {\em regularized\/} hamiltonian constraint as 
$
   \mathcal{H}_{T}[N]
     :=  \sum_{\Delta \in T} \mathcal{H}^{m}_{\Delta} [N] ~,
  $ 
with
 \begin{equation}
   \label{Hm_delta:classical2}
   \mathcal{H}^m_{\Delta}[N]
     := \frac{N(v)}{2 N^2_m} \, 
         \, \epsilon^{ijk} \,
        \mbox{Tr}\Big[h^{(m)}_{\alpha_{ij}} 
        h^{(m)}_{s_{k}} \big\{ 
        h^{(m)-1}_{s_{k}},V\big\}\Big] ~,
         \end{equation}
where $h^{(m)}_{\alpha_{ij}}$ and $h^{(m)}_{s_{k}}$ are the holonomies, in arbitrary representation $m$, around the triangle $\alpha_{ij}$ and along the segment $s_k$ respectively. This expression converges to the Hamiltonian constraint (\ref{Ham_T2}) if the triangulation is sufficiently fine \cite{Gaul:2000ba}. The expression (\ref{Hm_delta:classical2}) can finally  be promoted to a quantum operator,  since volume and holonomy have corresponding well-defined operators in LQG.  The lattice spacing of the triangulation $T$ then acts as a regularization parameter. 
Acting on a spin network state, the operator reduces to a sum over terms each acting on individual nodes. 
The continuum limit of the
operator turns out to be trivial in the quantum theory.  On 
diffeo\-mor\-phism-invariant (bra) states the regulator dependence drops out trivially. 
The result \cite{Gaul:2000ba, me1}, in the action of the operator ${\mathcal{H}}^m_{\Delta}$ on trivalent nodes, $| v(j_i,j_j,j_k)\rangle \equiv | v_3\rangle$, whereas $j_i,j_j,j_k$ are the
spins of the adjacent edges $e_i,\, e_j,\, e_k$, is given by: \be
   {\mathcal{H}}^m_{\Delta} \, \big| v (j_i,j_j,j_k) \big\rangle 
    = \: \frac{i l_0}{12 C(m)} \, 
     [ \; \sum_{a,b} 
A^{(m)}(j_i,a|j_j,b|j_k  )   
     \begin{array}{c} 
\ifx\JPicScale\undefined\def\JPicScale{0.3}\fi
\psset{unit=\JPicScale mm}
\psset{linewidth=0.3,dotsep=1,hatchwidth=0.3,hatchsep=1.5,shadowsize=1,dimen=middle}
\psset{dotsize=0.7 2.5,dotscale=1 1,fillcolor=black}
\psset{arrowsize=1 2,arrowlength=1,arrowinset=0.25,tbarsize=0.7 5,bracketlength=0.15,rbracketlength=0.15}
\begin{pspicture}(0,0)(90,80)
\psline[linewidth=0.2,linestyle=dashed,dash=0.5 0.5](30,37)(43,50)
\psline[linewidth=0.2,linestyle=dashed,dash=0.5 0.5](57,50)(70,37)
\psline[linewidth=0.2,linestyle=dashed,dash=0.5 0.5](57,77)(57,68)
\psline[linewidth=0.2,linestyle=dashed,dash=0.5 0.5](43,77)(43,58)
\psline[linewidth=0.2,linestyle=dashed,dash=0.5 0.5](50,38)(60,27)
\psline(50,47)(50,77)
\rput(17,17){$\scr{j_i}$}
\rput(83,17){$\scr{j_j}$}
\rput(50,80){$\scr{j_k}$}
\rput(39,32){$\scr{a}$}
\psline[linewidth=0.2,linestyle=dashed,dash=0.5 0.5](33,20)(30,17)
\psline(50,47)(80,17)
\psline[dotsize=1.3 2.5]{-*}(20,17)(50,47)
%\rput(33,52){$\scr{m}$}
\psline[linewidth=0.2,linestyle=dashed,dash=0.5 0.5](57,68)(57,50)
\psline[linewidth=0.2,linestyle=dashed,dash=0.5 0.5](70,37)(90,17)
\psline[linewidth=0.2,linestyle=dashed,dash=0.5 0.5](43,58)(43,50)
\psline[linewidth=0.2,linestyle=dashed,dash=0.5 0.5](10,17)(30,37)
\psline[linestyle=dotted](40,27)(60,27)
\psline[linestyle=dotted](33,20)(67,20)
\psline(73,24)(27,24)
\rput(50,22){$\scr{m}$}
\psline[linewidth=0.2,linestyle=dashed,dash=0.5 0.5](50,38)(39,27)
\psline[linewidth=0.2,linestyle=dashed,dash=0.5 0.5](66,20)(70,17)
\rput(62,32){$\scr{b}$}
\end{pspicture}
       \end{array}  
        + Permutations \; ]
        ~.
\label{fine thomas}
\ee
${\mathcal{H}}^m_{\Delta}| v_3\rangle$ gives the original state with the new "extraordinary link" $m$  between all the possible pairs of edges adjacent to the  node, with amplitudes \cite{me1} given by cyclic permutations of arguments. 

b) {\it A new operator} \cite{me1} is obtained if  we consider a regularization based not only on triangulations $T$ but also on the one-skeleton $\Gamma$, dual of $T$: $\Gamma$ is a graph with nodes $v$ in the center of the tetrahedra $v$ of $T$, and straight links that cut the triangles of $T$. In this case we fix a tetrahedron $v$ and one of its vertices, say $s$. Let $s^a$ be the segment that joins the center of $v$ to $s$, and $u^a$ and $v^a$ two of the sides of the triangle $s$ opposite to the tetrahedron's vertex $s$. The volume of the tetrahedron can be written as $V=\sum_s \frac 1{18}\epsilon_{abc}s^a u^b v^c$ where the sum is over the four vertices of the tetrahedron. Consider now the quantity 
\be
H_{\Delta}= \sum_s \ h^i_{\triangolo s}\ {\rm Tr}[\tau^i\, h_s^{-1}\{h_s, V\}]
 \label{alternativa1}
\ee
where $\triangolo\,s$ is the triangle opposite to the vertex $s$. 
If $A$ and $e$ are constant on the tetrahedron, it is easy to see (for instance using coordinates in which the tetrahedron is regular) that this gives 
\be
H_{\Delta}= \sum_s \  F^i_{ab}u^a v^b   s^c e_c^i=18 \;{\rm Tr}(F_{ab}e_c)\epsilon^{abc}V=18 \; \int_v {\rm Tr}(F\wedge e). 
 \ee
Therefore we can replace (\ref{Hm_delta:classical2}) with 
\be
\begin{split}
      \mathcal{H}_{\Delta}[N]
     :&= \frac{N(v)}{36 N_m^2} \, 
         \sum_s \,\ h^i_{\triangolo s}\ 
        {\rm Tr}\Big[\tau^i\, h_{s} \big\{ 
        h^{-1}_{s},V\big\}\Big]
        \label{H_delta2}
\end{split}
\ee
 where all holomonies are taken in the representation $m$, where $\tau^i$ are the (anti-hermitian) generators of $su(2)$ and $N^2_m={\rm Tr}\!\left[\tau^i\tau^i\right]=-(2m+1)m(m+1)$.  Notice that the sum is over the four links emerging from $v$ in $T$ and $\triangolo\, s$ is a triangle that joins the three points sitting on the three other links emerging from $v$. Notice also that this triangle and the center $v$ define a tetrahedron, which we shall denote $\tetraedro\,s$. It is then natural to replace the triangle regularization of the curvature with the tetrahedral one defined in \cite{me1}: 
the idea is to substitute $h^i_\triangolo\!\!$ with a different object, denoted 
${h}^i_\tetraedro$, and which is defined by the {\em spin-network function} of the connection $A$, associated to the tetrahedron generated by {\em three} segments $s_{01},s_{02},s_{03}$, emerging from a point $n$, with one open link and explicitly defined as  
\begin{equation}
	 {h}^{i}_\tetraedro
= c_m N_m\ 
i^{i\alpha\beta\gamma}\; i^{\delta\epsilon\zeta}\; i^{\theta\iota\kappa}\; i^{\lambda\mu\nu}\; \;h[{s_{01}}]_{\alpha\delta}\; h[s_{02}]_{\beta\kappa}\; h[s_{03}]_{\gamma\mu}\; h[s_{12}]_{\epsilon\theta}\; h[s_{23}]_{\iota\lambda}\; h[s_{31}]_{\nu\zeta}\; .
\end{equation}
Here $h$ are holomies in a representation of (integer) spin $m$. $i^{\alpha\beta\gamma}$ is the (unique) normalized 3-valent intertwiner between three representations $m$, and $i^{i\alpha\beta\gamma}$ is a normalized 4-valent intertwiner, with the first index in the adjoint representation and the other three in the representation $m$, satisfying $i^{i\alpha\beta\gamma}\;=i^{i\gamma\alpha\beta}\;=i^{i\beta\gamma\alpha}\;$ and $c^{-1}_m = i^{i\alpha\beta\gamma}\; i^{\alpha\epsilon\zeta}\; i^{\epsilon\beta\iota}\; i^{\lambda\gamma\zeta}\;i^{i\lambda\iota}$.
The key property of $h^i_\tetraedro$ is that in the limit in which the size of the tetrahedron is small, we have 
$
h^i_\tetraedro =  h^i_\triangolo
$
where the triangle is the face of the tetrahedron opposite to the 4-valent node $n$.

With this observation we can replace $h^i_\triangolo$ with $h^i_\tetraedro$ in (\ref{H_delta2}) and promote it to the quantum operator:
\be
\begin{split}
      \mathcal{\hat{H}_{\Delta}}[N]
      :&= \frac{-i}{36\cdot 8\pi \gamma l^2_p} \, 
    \sum_v
        \frac{N(v)}{N_m^2} \, \sum_s \,\ \hat{h}^i_{\tetraedro s}\ 
        {\rm Tr}\Big[\tau^i \hat{h}_{s} 
        \hat{V} \hat{h}^{-1}_{s} \Big]
        \label{H_delta4}
\end{split}
\ee
The action of the operator on a spin network state with support on a graph $\gamma$ can then be defined, following  \cite{Thiemann96a}, by choosing a regularizing triangulation $T$ adapted to $\gamma$. Here one has to choose $T$ such that $\gamma$ is a subgraph of $T^*$. 
To analyze the new constraint we restrict our attention to 4-valent nodes.
On a single 4-valent node, it is a sum over the four edges that emerge from the node, $H_{\Delta} \left|v_4\right\rangle=\sum_s\  H^s_{\Delta} \left|v_4\right\rangle$
with \be
\begin{split}
&	 \hat{H}^{s}_{\Delta}|v_4\rangle
=
%	        \\&=
N_m^2 c_m \!\!\!\!
\sum_{c,d,e,f,g,h,k} \,d_c \,d_f\,d_g\,d_h\,d_k\, \frac{l_0^3}{4} \,  
                \, V{}_{j_l,i_n}{}^{d,e} (j_i,j_j,j_k,m,c) \nonumber \\& \hspace{5em}\times \left\{
\begin{matrix}
                      j_l      & d & 1 \\
                      m      & m   & c                  
\end{matrix}
\right\} \ \ 
F_{\times}(m,d,e,f,g,h,k,j_i,j_j,j_k,j_l) \;  
           \hspace{-3em}     \begin{array}{c}
\ifx\JPicScale\undefined\def\JPicScale{0.4}\fi
\psset{unit=\JPicScale mm}
\psset{linewidth=0.3,dotsep=1,hatchwidth=0.3,hatchsep=1.5,shadowsize=1,dimen=middle}
\psset{dotsize=0.7 2.5,dotscale=1 1,fillcolor=black}
\psset{arrowsize=1 2,arrowlength=1,arrowinset=0.25,tbarsize=0.7 5,bracketlength=0.15,rbracketlength=0.15}
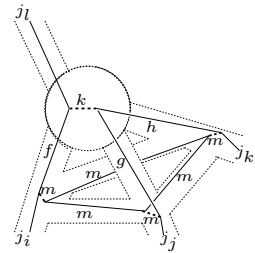
\begin{pspicture}(0,0)(102,78)
\psline[linewidth=0.2,linestyle=dashed,dash=0.5 0.5](26,5)(37,40)
\psline[linewidth=0.2,linestyle=dashed,dash=0.5 0.5](77,11)(80,4)
\psline[linewidth=0.2,linestyle=dashed,dash=0.5 0.5](37,8)(36,5)
\psline[linewidth=0.2,linestyle=dashed,dash=0.5 0.5](55,33)(64,16)
\psline(34,18)(44,46)
\psline[linewidth=0.2,linestyle=dashed,dash=0.5 0.5](62,39)(70,36)
\psline[linewidth=0.2,linestyle=dashed,dash=0.5 0.5](84,32)(73,19)
\psline[linewidth=0.2,linestyle=dashed,dash=0.5 0.5](93,30)(77,11)
\psline[linewidth=0.2,linestyle=dashed,dash=0.5 0.5](56,30)(42,25)
\psline[linewidth=0.2,linestyle=dashed,dash=0.5 0.5](59,23)(45,18)
\psline[linewidth=0.2,linestyle=dashed,dash=0.5 0.5](70,36)(65,34)
\psline[linewidth=0.2,linestyle=dashed,dash=0.5 0.5](82,33)(68,28)
\psline[linewidth=0.2,linestyle=dashed,dash=0.5 0.5](64,16)(44,16)
\psline[linewidth=0.2,linestyle=dashed,dash=0.5 0.5](68,8)(37,8)
\psline[linewidth=0.2,linestyle=dashed,dash=0.5 0.5](36,75)(44,58)
\psline[linewidth=0.2,linestyle=dashed,dash=0.5 0.5](26,75)(37,52)
\psline(44,46)(31,75)
\psline[linewidth=0.2,linestyle=dashed,dash=0.5 0.5](64,48)(101,37)
\rput{90}(50.2,45.54){\psellipticarc[linewidth=0.2,linestyle=dashed,dash=0.5 0.5](0,0)(14.5,-14){-27.38}{80.22}}
\rput{0}(50,46){\psellipticarc[linewidth=0.2,linestyle=dashed,dash=0.5 0.5](0,0)(14,-14){67.38}{111.04}}
\rput{50.55}(50.1,46.02){\psellipticarc[linewidth=0.2,linestyle=dashed,dash=0.5 0.5](0,0)(14.06,-13.96){-151.8}{-104.8}}
\rput(30,58){}
\rput(29,78){$\scr{j_l}$}
\rput(48,49){$\scr{k}$}
\psline[linewidth=0.2,linestyle=dashed,dash=0.5 0.5](45,33)(42,25)
\psline[linewidth=0.2,linestyle=dashed,dash=0.5 0.5](68,8)(70,4)
\psline[linewidth=0.2,linestyle=dashed,dash=0.5 0.5](62,39)(73,18)
\psline[linewidth=0.2,linestyle=dashed,dash=0.5 0.5](92,30)(98,28)
\psline[linewidth=0.2,linestyle=dashed,dash=0.5 0.5](82,33)(84,32)
\psline(69,12)(36,15)
\psline(59,25)(36,15)
\psline(90,37)(68,29)
\psline(90,37)(69,12)
\rput(29,3){$\scr{j_i}$}
\rput(77,2){$\scr{j_j}$}
\rput(102,31){$\scr{j_k}$}
\psline[border=0.5](94,38)(100,32)
\psline[border=0.5](53,46)(74,10)
\psline[linewidth=0.6,linestyle=dashed,dash=1 1,border=0.4](44,46)(53,46)
\rput(49,11){$\scr{m}$}
\rput(82,23){$\scr{m}$}
\rput(52,24){$\scr{m}$}
\psline(31,5)(34,18)
\rput(37,32){$\scr{f}$}
\psline[border=0.5](74,10)(75,4)
\psline[border=0.5](53,46)(94,38)
\rput(61,28){$\scr{g}$}
\rput(71,40){$\scr{h}$}
\psline[linewidth=0.6,linestyle=dashed,dash=1 1](36,15)(34,18)
\psline[linewidth=0.6,linestyle=dashed,dash=1 1](68,12)(74,10)
\psline[linewidth=0.6,linestyle=dashed,dash=1 1](94,38)(90,37)
\psline[linewidth=0.2,linestyle=dashed,dash=0.5 0.5](45,18)(44,16)
\rput(71,9){$\scr{m}$}
\rput(37,19){$\scr{m}$}
\rput(92,35){$\scr{m}$}
\end{pspicture}
\end{array}
\end{split}
\label{finale}
\ee 
where $V{}_{j_l,i_n}{}^{d,e}$ is the volume of a non-gauge-invariant 5-valent node, the expression for F is given in \cite{me1} and remarkably contains 15j symbols.

\section{Conclusions}
The main differences between the operator (\ref{H_delta4}) and the quantum version of \eqref{H_delta2} are: 1)
The curvature is computed on a surface that is properly \emph{dual} to the direction indicated by $e$. Notice in fact that it is computed on a triangle that \emph{surrounds} the direction of the link $s$. This is different from the old case; there, $\gamma$ had to be a subgraph of the triangulation $T$ itself, not its dual. 
2) The new operator creates {\em three} new links instead than one. Therefore generically it transforms a 4-valent node into \emph{four} nodes, thus implementing the 1-4 Pachner move \cite{lqgcov}. This is precisely the action of the dynamics in the simplicial spinfoam models. 
3) The nodes created are themselves 4-valent. Thus are ``of the same kind" as the original node and therefore with volume.  This is not the case as the old operator  
4) Finally, when we compute matrix elements of this operator, we find $15j$ Wigner symbols, even if a way to recover from these one of the known spinfoam model is still under investigation.

Thiemann's proof that the Hamiltonian operator is anomaly free  \cite{Thiemann96a} does not go through with the new operator.  The proof was indeed based on the fact that the nodes generated were rather ``special" and had no volume, and therefore the Hamiltonian could not act again on them. On the other hand, this fact is sometime viewed as one of the unconvincing aspect \cite{dubbi} of the old construction, and this is partially corrected with the constraint considered here. (On different ways to address the issue, see in particular \cite{Giesel:2006uj,master}.)


\section*{References}

\begin{thebibliography}{99}

\bibitem{lqgcan}

A.\,Ashtekar and J.\,Lewandowski, 2004
%``Background independent quantum gravity: A
  %status report'', 
  {\it Class.\,Quant.\,Grav.} {\bf 21}  R53
   \nonum
T.\,Thiemann, 2007 {\it Modern canonical quantum general relativity} (Cambridge: Cambridge
  University Press)
  \nonum
C.\,Rovelli, 2004 {\it Quantum Gravity} (Cambridge: Cambridge
  University Press).

\bibitem{lqgcov}
 A.~Perez, 2003 
 {\it Class.\ Quant.\ Grav.\ }  {\bf 20} R43 

%\bibitem{BarrettCrane98} 
  %J.~W.~Barrett and L.~Crane,
  %``Relativistic spin networks and quantum gravity,''
  %J.\ Math.\ Phys.\  {\bf 39}, 3296 (1998)
  %[arXiv:gr-qc/9709028].
  %%CITATION = JMAPA,39,3296;%%

  
\bibitem{lqg}
   C.~Rovelli, 2011
 {\it Class.\ Quant.\ Grav.\ } {\bf 28} 114005 
   %%CITATION = ARXIV:1004.1780;%

\bibitem{scattering} 
  L.~Modesto and C.~Rovelli, 2005
 {\it Phys.\ Rev.\ Lett.\ } {\bf 95} 191301 
  %%CITATION = PRLTA,95,191301;%%
\nonum
 E.~Alesci and C.~Rovelli, 2007
{\it  Phys.\ Rev.\  D} {\bf 76} 104012
    %%CITATION = PHRVA,D76,104012;%%
\nonum
E.~Alesci and C.~Rovelli, 2008
 {\it Phys.\ Rev.\  D} {\bf 77} 044024 
    %%CITATION = PHRVA,D77,044024;%%
\nonum
E.~Bianchi, E.~Magliaro and C.~Perini, 2009
 {\it Nucl.\ Phys.\  B} {\bf 822}  245 

\bibitem{ReisenbergerRovelli97}  
  M.~P.~Reisenberger and C.~Rovelli, 1997
 {\it Phys.\ Rev.\  D} {\bf 56} 3490 
    %%CITATION = PHRVA,D56,3490;%%
 
 \bibitem{Io karim e francesco}
  E.~Alesci, K.~Noui and F.~Sardelli, 2008
{\it  Phys.\ Rev.\  D} {\bf 78} 104009 

 \bibitem{me2}
  E.~Alesci, T.~Thiemann, A.~Zipfel, 2001
 Linking covariant and canonical LQG: New solutions to the Euclidean Scalar Constraint 
  {\it Preprint} arXiv:1109.1290
 
    
    \bibitem{Noui Perez}
K.~Noui and A.~Perez, 2005
{\it Class.\ Quant.\ Grav.\ } {\bf 22} 1739   
                 
\bibitem{Engle:2007uq}
  J.~Engle, E.~Livine, R.~Pereira and C.~Rovelli, 2008
{\it  Nucl.\ Phys.\  B} {\bf 799} 136
    %%CITATION = NUPHA,B799,136;%%
    \nonum

 L.~Freidel and K.~Krasnov, 2008
{\it  Class.\ Quant.\ Grav.\ } {\bf 25} 125018
    %%CITATION = CQGRD,25,125018;%%
  
\bibitem{Kaminski:2009fm}
W.~Kaminski, M.~Kisielowski and J.~Lewandowski, 2010
{\it  Class.\ Quant.\ Grav.\  } {\bf 27} 095006 
    %%CITATION = CQGRD,27,095006;%%

    


\bibitem{Thiemann96a}  T.~Thiemann, 1998
 {\it Class.\ Quant.\ Grav.\ }  {\bf 15}  839
  %%CITATION = CQGRD,15,839;%%

  
\bibitem{Thiemann96b} T.~Thiemann, 1998
{\it  Class.\ Quant.\ Grav.\ }  {\bf 15} 875 
  %%CITATION = CQGRD,15,875;%%
  
\bibitem{AshtekarLewand98}  A.~Ashtekar and J.~Lewandowski, 1998
  {\it Adv.\ Theor.\ Math.\ Phys.\ }  {\bf 1} 388 
  %%CITATION = 00203,1,388;%%  
  
\bibitem{me1}
E.~Alesci and C.~Rovelli,  2010
{\it Phys.\ Rev.\ D } {\bf 82}
044007

\bibitem{Gaul:2000ba}
  M.~Gaul and C.~Rovelli, 2001
{\it  Class.\ Quant.\ Grav.\ }  {\bf 18}  1593

\bibitem{dubbi} R.~Gambini, J.~Lewandowski, D.~Marolf and J.~Pullin, 1998
 {\it Int.\ J.\ Mod.\ Phys.\  D} {\bf 7} 97 
  %%CITATION = IMPAE,D7,97;%%
\nonum
J.~Lewandowski and D.~Marolf, 1998
   {\it Int.\ J.\ Mod.\ Phys.\  D} {\bf 7} 299 
  \nonum
 L.~Smolin, 96 
   The classical limit and the form of the Hamiltonian constraint in
  non-perturbative quantum general relativity
  {\it Preprint}  arXiv:gr-qc/9609034.
  
\bibitem{Giesel:2006uj}
  K.~Giesel and T.~Thiemann, 2007
{\it Class.\ Quant.\ Grav.\ } {\bf 24} 2465 

\bibitem{master}
  T.~Thiemann, 2006
  {\it Class.\ Quant.\ Grav.\ } {\bf 23}, 2211 
\nonum
B.~Dittrich and T.~Thiemann, 2006
  {\it Class.\ Quant.\ Grav.\ } {\bf 23}, 1025 
  %%CITATION = CQGRD,23,1025;%%
 \nonum 
  T.~Thiemann, 2006
  {\it Class.\ Quant.\ Grav.\ }  {\bf 23} 2249 
  

\end{thebibliography}
\end{document}